\documentclass[a4paper,12pt]{article}

\usepackage{graphicx} \usepackage{amsmath,amssymb,cite,comment}

\usepackage{amsfonts}

\def\idos{invariant differential operators}
\def\Idos{Invariant differential operators}
\def\ido{invariant differential operator}

\def\llr{\longrightarrow}
  \def\tV{{\tilde V}}

 \def\cgc{{\cg^\bbc}}
 \def\np{\vfill\eject}
\def\a{\alpha}
\def\b{\beta}
\def\d{\delta}

\def\vr{\vert}

\def\D{{\Delta}}
\def\ca{{\cal A}} \def\cb{{\cal B}} \def\cc{{\cal C}}
\def\cd{{\cal D}}  \def\cf{{\cal F}}
\def\cg{{\cal G}} \def\ch{{\cal H}} 
 \def\ck{{\cal K}} 
\def\cm{{\cal M}} \def\cn{{\cal N}} 
\def\cp{{\cal P}} \def\cq{{\cal Q}}  
\def\ct{{\cal T}}

\def\bbz{\mathbb{Z}}
\def\bbc{\mathbb{C}}
\def\bbr{\mathbb{R}}
\def\bbn{\mathbb{N}}

\def\eqn#1{\begin{equation}\label{#1}}
\def\ee{\end{equation}}

\def\bea{\begin{eqnarray}}
\def\eea{\end{eqnarray}}

\def\eqnn#1{\begin{eqnarray}\label{#1}}
\newcommand{\eqna}[1]{\begin{subequations} \label{#1}
\begin{eqnarray}}
\def\eena{\end{eqnarray}
\end{subequations}}

\def\nn{\nonumber}

\def\d{\delta}
\def\L{\Lambda}

   \def\nt{\noindent}

\def\veps{\varepsilon}
\def\r{{\rho}}

\input epsf.tex
\newcount\figno
\def\fig#1#2#3{
\par\begingroup\parindent=0pt\leftskip=1cm\rightskip=1cm\parindent=0pt
\baselineskip=11pt \global\advance\figno by 1 
\epsfxsize=#3 \centerline{\epsfbox{#2}} \vskip 12pt
#1\par
\endgroup\par}
\def\figlabel#1{\xdef#1{\the\figno}}
\def\encadremath#1{\vbox{\hrule\hbox{\vrule\kern8pt\vbox{\kern8pt
\hbox{$\displaystyle #1$}\kern8pt} \kern8pt\vrule}\hrule}}

\font\tfont=cmbx12 scaled\magstep1 

\begin{document}

\begin{center}

{\tfont Invariant Differential Operators for the Real Exceptional Lie Algebra $F'_4$}

\vskip 1.5cm

{\bf V.K. Dobrev}

 \vskip 5mm


{Institute of Nuclear Research and Nuclear Energy,\\
Bulgarian Academy of Sciences,\\ 72 Tsarigradsko Chaussee, 1784
Sofia, Bulgaria}

 \end{center}

\vskip 1.5cm

 \centerline{{\bf Abstract}}
 
 {In the present paper we continue the project of systematic
construction of invariant differential operators on the example of
the non-compact  exceptional Lie algebra  $F'_4$ which is split real form of the exceptional Lie algebra $F_4$.
We consider induction from a maximal parabolic algebra. We classify
the reducible Verma modules over $F_4$ which are compatible with this induction.
Thus, we obtain the classification of  the corresponding invariant differential operators.}

\vskip 1.5cm

{\it Presented at Corfu Summer Institute 2019 "School and Workshops on Elementary Particle Physics and Gravity" (CORFU2019)\\
		31 August - 25 September 2019, 
	Corfù, Greece\\ 
Proceedings, Volume 376, PoS (CORFU2019)    (Published on: August 18, 2020)  233.}

\vskip 1.5cm









 \pagestyle{empty}

\section{Introduction}

Invariant differential operators   play very important role in the
description of physical symmetries. The general scheme for constructing these
operators was given some time ago \cite{Dob}.
In recent papers \cite{Dobinv,Dobparab} we started the systematic explicit
construction of the invariant differential operators.

The first task in the construction is to make the multiplet classification of
the reducible Verma modules over the algebra in consideration following \cite{Dobmul}.
Such classification provides the weights of embeddings between the Verma modules via
the singular vectors, and thus, by \cite{Dob}, the weights of the invariant differential operators.

We have done the multiplet classification for many real non-compact algebras, first from the class
of algebras that have discrete series representations, see \cite{Dobk1}.
In the present paper we  focus on the complex exceptional Lie algebra ~$F_4$~ and on its
 split real form algebra  $F'_4$. Our scheme requires that we use induction from parabolic subalgebras.
 In the present paper we choose a parabolic subalgebra containing the factor ~$\cm \oplus \ca$, where
 ~$\cm ~=~ sl(3,\bbr) \oplus sl(2,\bbr)$, ~$\ca ~=~ o(2)$.
 This choice is motivated by the fact that the complexification of $\cm\oplus \ca$ and the corresponding compact form
 ~$\cm_c\oplus \ca_c ~=~ su(3) \oplus su(2)\oplus u(1)$~ have applications in physics being the Lie algebra symmetry of the standard
 model of elementary particles \cite{Wein} (see also \cite{ITT}).\footnote{More precisely the symmetry is presented on the group level as :
 ~$ G=SU(3) \times SU(2) \times U(1)/Z$, where ~$Z$ belongs to the center of $G$.}

  We present the multiplet classification of
 the reducible Verma modules over $F_4$ which are compatible with the chosen parabolic of $F'_4$.
 We give also the weights of the singular vectors between
 these modules. By the scheme of \cite{Dob} these  singular vectors will produce
 the invariant differential operators.

\section{Preliminaries}

\subsection{Lie algebra}

We start with the complex exceptional Lie algebra  ~$\cg^\bbc ~=~ F_4$. We use
the standard definition of ~$\cg^\bbc$~ given in terms of the
Chevalley generators $X^\pm_i ~, ~H_i ~, ~i=1,2,3,4 (=$rank$\,F_4)$, by the relations~:
\eqnn{com}
 &[H_j\,, \,H_k] \, = \, 0 \,, \,\,\,[H_j\,, \,X^\pm_k] \, = \, \pm
a_{jk} X^\pm_k \,,
\,\,\,[X^+_j \,, \,X^-_k] \, = \,
\d_{jk} \,H_j \,,  \\
&\sum_{m=0}^n \,(-1)^m \,\left({n \atop m}\right)
\,\left(X^\pm_j\right)^m \,X^\pm_k \,\left(X^\pm_j\right)^{n-m}
\,=\, 0 \,, \,\,j \neq k \,, \,\,n = 1 - a_{jk} \,, \nn\eea where
\eqn{aijf4} (a_{ij}) = \begin{pmatrix} 2 & -1 & 0 & 0 \cr -1 & 2
& -1 & 0\cr 0 & -2 & 2 & -1\cr 0 &0 &-1 & 2\end{pmatrix} ~; \ee
is the Cartan matrix of $\cg^\bbc$, ~$\a^\vee_j
\,\equiv\, {2 \a_j\over (\a_j , \a_j)}$ is the co-root of
$\a_j\,$, \,\, $(\cdot , \cdot)$ is the scalar product of the roots,
so that the nonzero products between the simple roots are:
$(\a_1, \a_1) = (\a_2, \a_2) = 2(\a_3, \a_3) = 2(\a_4, \a_4)~=~2 $, ~
$(\a_1, \a_2) = -1$, ~$(\a_2, \a_3) = -1$, ~$(\a_3, \a_4) = -1/2$.
 The
elements $H_i$ span the Cartan subalgebra $\ch$ of $\cg^\bbc$, while
the elements $X^\pm_i$ generate the subalgebras $\cg^\pm$. We shall
use the standard triangular decomposition \eqn{deca} \cg^\bbc \,=\,
\cg_+\oplus\ch\oplus \cg_- \,, \qquad\cg_\pm \,\equiv
\,\mathop{\oplus}\limits_{\a\in\D^\pm} \,\cg_\a \,, \ee where
$\D^+$, $\D^-$, are the sets of positive, negative, roots, resp.
Explicitly    we have that there are roots of two lengths with
length ratio $2:1$.\\ The long roots are:
~$\a_1$, ~$\a_2$,
~$\a_1 + \a_2$,
~$\a_2 + 2\a_3$,
~$\a_1 + \a_2 + 2\a_3$,
~$\a_1 + 2\a_2 + 2\a_3$,
~$\a_2 +2\a_3 + 2\a_4$,
~$\a_1 +\a_2 +2\a_3 +2\a_4$,
~$\a_1 +2\a_2 +2\a_3 +2\a_4$,
~$\a_1 +2\a_2 + 4\a_3 +2\a_4$,
~$\a_1 + 3\a_2 + 4\a_3 + 2\a_4$,
~$2\a_1 + 3\a_2 + 4\a_3 + 2\a_4$.
With the chosen normalization they have length 2.\\
The short roots are:
~$\a_3$, ~$\a_4$,
~$\a_2 + \a_3$,
~$\a_3 + \a_4$,
~$\a_1 + \a_2 + \a_3$,
~$\a_2 + \a_3 + \a_4$,
~$\a_1 +\a_2 +\a_3 +\a_4$,
~$\a_2 +2\a_3 +\a_4$,
~$\a_1 +2\a_2 +2\a_3+\a_4$,
~$\a_1 +\a_2 +2\a_3 +\a_4$,
~$\a_1 +2\a_2 + 3\a_3 + \a_4$,
~$\a_1 +2\a_2 +3\a_3 +2\a_4$,
and they have length 1.\\
(Note that the short roots are exactly those which contain $\a_3$
and/or $\a_4$ with odd coefficient, while the long roots contain $\a_3$
and $\a_4$ with even coefficients.)

Thus, $F_4$ is 52--dimensional ($52 = \vert \D\vert +$ rank $F_4$).

In terms of the normalized basis ~$\veps_1, \veps_2, \veps_3,
\veps_4$ ~we have:
\eqnn{posf4} \D^+ = \{\veps_i, ~1 \leq i \leq 4; ~\veps_j \pm
\veps_k, ~1 \leq j <k \leq 4; \nn\\ {1 \over 2}(\veps_1 \pm \veps_2 \pm
\veps_3\pm \veps_4), {\rm ~all\ signs}\} ~. \eea The simple roots are: \eqn{pif4}
\pi = \{\a_1 = \veps_2 - \veps_3, ~\a_2 = \veps_3 - \veps_4, \a_3 = \veps_4,
~\a_4 = {1 \over 2} (\veps_1 - \veps_2 - \veps_3 - \veps_4)\}
~.\ee

The maximal compact subalgebra is ~$\ck ~=~ sp(3)\oplus su(2)$. Its complexification ~$\ck^\bbc$~ may be embedded most easily in ~$F_4$~
as the Lie algebra generated by the subalgebras with simple roots ~$\{ \a_2,\a_3,\a_4\}$~ and  ~$\{ \a_1\}$.
The long roots of $sp(3,\bbc)$ in this embedding are:
 ~$\a_2$, ~$\a_2 + 2\a_3$, ~$\a_2 +2\a_3 + 2\a_4$.
 The short roots are:
~$\a_3$, ~$\a_4$,
~$\a_2 + \a_3$,
~$\a_3 + \a_4$,
 ~$\a_2 + \a_3 + \a_4$,
 ~$\a_2 +2\a_3 +\a_4$.

Note that the 16 roots on the first line of \eqref{posf4} form the positive root system
of ~$B_4$~ with simple roots ~$\veps_i-\veps_{i+1}\,, ~i=1,2,3, ~\veps_4\,$.

The Weyl group of ~$F_4$~ is the semidirect product of ~$S_3$~ with
a group which itself is the semidirect product of ~$S_4$~ with
~$(\bbz/2\bbz)^3$, thus, ~$|W| = 2^7\,3^2 = 1152$.

\subsection{Verma modules}

Let us recall that a ~{\it Verma module} ~$V^\L$~ is defined as
the HWM over ~$\cgc$~ with highest weight ~$\L \in \ch^*$~ and
highest weight vector ~$v_0 \in V^\L$, induced from the
one-dimensional representation ~$V_0 \cong \bbc v_0$~ of
~$U(\cb)$~, where ~$\cb  = \ch \oplus \cg_+$~ is a Borel
subalgebra of ~$\cgc$, such that:
\eqnn{indb}
& &X ~v_0 ~~=~~ 0 , \quad \forall\, X\in \cg_+ \cr
&&H ~v_0 ~~=~~ \L(H)~v_0\,, \quad \forall\, H \in \ch \eea

Verma modules are generically irreducible. A Verma  module ~$V^\L$~ is
reducible \cite{BGG} iff there exists a root ~$\b \in\D^+$~ and ~$m\in\bbn$~
such that
\eqn{red} (\L + \r~, ~\b^\vee) ~=~  m \ee
holds, where ~$\r = {1 \over 2}\sum_{\a \in \D^+}~\a$~,
($ \r ~=~ 8\a_1 +15\a_2 +21\a_3 + 11\a_4 $).

If \eqref{red} holds then the reducible Verma module $V^\L$  contains an invariant submodule
which is also a Verma module ~$V^{\L'}$~ with shifted weight ~$\L'=\L-m\b$.
This statement is equivalent to the fact that $V^\L$ contains a
~{\it singular vector}~
~$v_s \in V^\L$, such that ~$v_s ~\neq ~\xi v_0\,$, ($0\neq\xi\in\bbc$),
and~:
\eqnn{sing}
&& X ~v_s ~~=~~ 0 , \quad \forall\, X\in \cg_+ \cr
&&H ~v_s ~~=~~ \L'(H) ~v_s\,, \quad
\L' ~=~ \L - m\b, ~~\forall\,
H \in \ch \eea
More explicitly, \cite{Dob},
\eqn{singp} v^s_{m,\b} = \cp_{m,\b}\, v_0 \ . \ee

   The general reducibility conditions \eqref{red} for $V^\L$
   spelled out for the  simple roots in our situation are:
\eqnn{reda}
&& m_1 ~\equiv~ m_{\a_1} ~=~ (\L +\r, \a_1), ~~~
m_2 ~\equiv~  m_{\a_2} ~=~ (\L +\r, \a_2),\\
&& m_3 ~\equiv~  m_{\a_3} ~=~ (\L +\r, 2\a_3), ~~~
m_4 ~\equiv~ m_{\a_4} ~=~ (\L +\r, 2\a_4) \nn\eea
The numbers ~$m_i$~ from \eqref{reda} corresponding to the simple roots are called Dynkin labels, while the more general Harish-Chandra parameters  are:
\eqn{hcpar} m_\b ~=~ (\L +\r, \b^\vee), ~~~\b\in\D^+ \ee

\subsection{Structure theory of the real form}

The split real form of ~$F_4$~ is denoted as ~$F'_4\,$, sometimes as  ~$F_{4(4)}\,$.
This real form has   discrete series representations since ~rank$F'_4\,~=$ ~rank$\,\ck$.  We can use
the same basis (but over $\bbr$) and the same root system.

The Iwasawa decomposition of the real split form ~$\cg \equiv F'_4\,$,~ is:
\eqn{iwa32} \cg ~=~ \ck \oplus \ca_0 \oplus \cn_0 \ , \ee
the Cartan decomposition is:
\eqn{car32} \cg ~=~ \ck \oplus \cq , \ee
where we use: the maximal compact subgroup ~$\ck \cong
sp(3) \oplus su(2)$, ~$\dim_\bbr\,\cq = 28$,
 ~$\dim_\bbr\,\ca_0 = 4$, ~$\cn_0 =  \cn^+_0\,$, or
 ~$\cn_0 =  \cn^-_0 \cong \cn^+_0\,$, ~$\dim_\bbr\,\cn_0^\pm =24$.

   Since ~$\cg$~ is maximally split, then
the centralizer ~$\cm_0$~ of ~$\ca_0$~ in ~$\ck$~ is zero, thus, the minimal parabolic ~$\cp_0$~
and the corresponding Bruhat decomposition are:
\eqn{min32} \cp_0 ~=~ \ca_0 \oplus \cn_0 \ , \qquad \cg ~=~ \ca_0 \oplus \cn^+_0\oplus \cn^-_0 \ee

The importance of the parabolic subgroups comes from the fact that
the representations induced from them generate all (admissible)
irreducible representations of the group under consideration \cite{Lan,Zhea,KnZu}.

We recall that in general a parabolic subalgebra ~$\cp ~=~ \cm \oplus \ca \oplus \cn$~ is any subalgebra of $\cg$
which contains a minimal parabolic subalgebra ~$\cm_0\,$. In general, ~$\cm$~ contains the subalgebra $\cm_0$, while
~$\ca$~ is contained in ~$\ca_0$, ~$\cn$~ is contained in ~$\cn_0$.\\
On the other extreme are the maximal parabolic subalgebras for which ~dim$\,\ca=1$.

\subsection{Elementary representations}

Further, let ~$G,K,P,M,A,N$~ are Lie groups with Lie algebras  ~$\cg_0,\ck,\cp,\cm,\ca,\cn$.

Let ~$\nu$~ be a (non-unitary) character of ~$A$, ~$\nu\in\ca^*$.
 Let ~ $\mu$ ~ fix a finite-dimensional (non-unitary) representation
~$D^\mu$~ of $M$ on the space ~$V_\mu\,$.
 In the case when $M$ is cuspidal then we may  use also  the
 discrete series representation of $M$ with the same Casimirs as $D^\mu$.
(We ignore a possible discrete center of $M$ since its representations are not
relevant for the construction of invariant differential operators \cite{Dobk1}.)

 We call the induced
representation ~$\chi =$ Ind$^G_{P}(\mu\otimes\nu \otimes 1)$~ an
~{\it \it elementary representation} of $G$ \cite{DMPPT}. (These are
called {\it generalized principal series representations} (or {\it
limits thereof}) in \cite{Knapp}.)   Their spaces of functions are:  \eqn{func}
\cc_\chi ~=~ \{ \cf \in C^\infty(G,V_\mu) ~ \vr ~ \cf (gman) ~=~
e^{-\nu(H)} \cdot D^\mu(m^{-1})\, \cf (g) \} \ee where ~$a=
\exp(H)\in A$, ~$H\in\ca\,$, ~$m\in M$, ~$n\in N$. The
representation action is the {\it left regular action}:  \eqn{lrega}
(\ct^\chi(g)\cf) (g') ~=~ \cf (g^{-1}g') ~, \quad g,g'\in G\ .\ee

An important ingredient in our considerations are the
highest/lowest weight representations~ of ~$\cg$. These can be
realized as (factor-modules of) Verma modules ~$V^\L$~ over
~$\cg$, where ~$\L\in (\ch)^*$,   the weight ~$\L = \L(\chi)$~ being determined
uniquely from $\chi$ \cite{Dob}.

As we have seen when a Verma module is reducible and \eqref{red} holds then there is a
singular vector \eqref{singp}. Relatedly, then
there exists \cite{Dob} an {\it \ido}  \eqn{invop}  \cd_{m,\b} ~:~ \cc_{\chi(\L)}
~\llr ~ \cc_{\chi(\L-m\b)} \ee given explicitly by: \eqn{singvv}
 \cd_{m,\b} ~=~ \cp_{m,\b}(\widehat{\cn^-})  \ee where
~$\widehat{\cn^-}$~ denotes the {\it right action} on the functions
~$\cf$.

Actually, since our ERs  are induced from finite-dimensional
representations of ~$\cm$~  the corresponding Verma modules are
always reducible. Thus, it is more convenient to use ~{\it
generalised Verma modules} ~$\tV^\L$~ such that the role of the
highest/lowest weight vector $v_0$ is taken by the
(finite-dimensional) space ~$V_\mu\,v_0\,$.

Algebraically, the above is governed by the notion of ~$\cm$-compact roots of $\cgc$. These are the roots of
$\cgc$ which can be identified as roots of ~$\cm^\bbc$~ as the latter  is a subalgebra of $\cgc$. The consequence
of this is that \eqref{red} is always fulfilled for the ~$\cm$-compact roots of $\cgc$. That is why we consider
generalised Verma modules. Relatedly, the \idos\ corresponding to  ~$\cm$-compact roots are trivial.

\def\dia{~~$\diamondsuit$}

\section{\Idos\ for $F'_4$}

The real form $F'_4$ has several parabolic subalgebras \cite{Dobinv}. We shall consider the maximal
parabolic subalgebra \cite{Dobinv}:
\eqnn{maxs} &&\cp ~=~ \cm \oplus \ca \oplus \cn  \ , \nn\\
&& \cm ~=~ sl(3,\bbr) \oplus \sl(2,\bbr) \ , \\
&& \dim\ca =1, \qquad \dim\cn = 20 \nn\eea
such that the embedding of $\cm$ and $\cm^\bbc$ in $\cgc$ is given by:
\eqn{embsr} sl(3,\bbr)^\bbc : \{ \a_1,\a_2,\a_{12}=\a_1+\a_2 \}, \qquad
sl(2,\bbr)^\bbc : \{ \a_4 \} \ee

\nt {\bf Remark:}~~
Note that $F'_4$ has a another maximal parabolic subalgebra that is also written as \eqref{maxs} but the
 embedding of $\cm$ and $\cm^\bbc$ flips the short and long roots \cite{Dobinv}:
 \eqn{embsrr} sl(3,\bbr)^\bbc : \{ \a_3,\a_4,\a_{34}=\a_3+\a_4 \}, \qquad
sl(2,\bbr)^\bbc : \{ \a_1 \} \ee
That case is also very interesting and will be considered next \cite{Dobf}. \dia

Further  we classify the generalized Verma modules (GVM) relative to the  maximal parabolic subalgebra \eqref{maxs}.
This also provides the classification of the $P$-induced ERs  with the same Casimirs.
 The classification is done as follows.   We group the
reducible Verma modules  (also the corresponding ERs) related by nontrivial embeddings
in sets called ~{\it multiplets} \cite{Dobmul,Dob}. These multiplets
may be depicted as a connected graph, the
vertices of which correspond to the GVMs  and the lines
between the vertices correspond to the GVM embeddings (and also the \idos\ between the ERs).
The explicit parametrization of the multiplets and of their Verma modules (and ERs) is
important for understanding of the situation.

The result of our classification is a follows. The multiplets of GVMs (and ERs) induced from
\eqref{maxs} are parametrized by four positive integers ~$\chi ~=~ [ m_1,m_2,m_3,m_4]$. Each
multiplet contains 96 GVMs (ERs). They are presented in  Fig. 1.

On the figure each arrow represents an embedding between two Verma modules, ~$V^\L$ and ~$V^{\L'}$,~ the arrow pointing to the embedded module ~$V^{\L'}$.
Each arrow carries a number ~$n$, $n=1,2,3,4$, which indicates the level of the embedding,  ~$\L' = \L - m_n\,\b$ \cite{Dobk1}.
Another feature is indicated by the enumeration of the GVMs (ERs). Namely,  if ~$\L$~ corresponds to signature ~$\chi^-_{k,\ell}$, ~$k<9$, then
~$\L'$~ corresponds to signature ~$\chi^-_{k+1,\ell'}$ (where ~$\ell,\ell'$~ are secondary enumerations that are absent in some cases).
Analogously:   if ~$\L$~ corresponds to signature ~$\chi^+_{k,\ell}$, ~$k>10$, then
~$\L'$~ corresponds to signature ~$\chi^+_{k-1,\ell'}$;  ~if ~$\L$~ corresponds to signature ~$\chi^-_{9,\ell}$,  then
~$\L'$~ corresponds to signature ~$\chi^*_{10,\ell'}$, (where $*$ may happen to be $'+'$ or $'-'$); ~if ~$\L$~ corresponds to signature ~$\chi^*_{10,\ell}$,  then
~$\L'$~ corresponds to signature ~$\chi^+_{9,\ell'}$.

Further, we can see there is an additional symmetry. 
It is relevant for the ERs and indicates the integral intertwining Knapp-Stein (KS) operators acting between the ERs.
Due to this symmetry in the actual parametrization   we shall use the conformal weight ~$d ~=~ 7/2 +c $, more precisely, the parameter $c$, instead of the non-compact Dynkin  label $m_3$.
 The parameter ~$c$~ is more convenient since the KS operators
flip its sign. The KS operators also involve an ~$sl(3)$~ flip of the Dynkin labels $m_1,m_2$ (see below).
Thus, the entries are:
\eqn{chihh} \chi ~=~ \{ n_1, n_2, c, n_4\} \ee
so that for the top ER (GVM) on the figure ~$\L^-_0$~ we have:
\eqn{chihhh} \chi^-_0 ~=~ \{ n_1=m_1, n_2=m_2, c = -(m_1+m_2+m_3+m_4/2), n_4=m_4\} \ee

Furthermore the ~$sl(3)$~ flip ~$( n_1, n_2)^\pm$ will be given below by:
 \eqn{sl3f}  ( n_1, n_2)^+ ~=~ ( n_1, n_2), \qquad ( n_1, n_2)^- ~=~ ( n_2, n_1)\ee

  Altogether, the explicit parametrization of the multiplets is given by:
  \eqnn{tabl}
\chi^\mp_0 ~&=&~ \{\, (m_1, m_2)^\pm, \mp  (m_1+m_2+m_3 + m_4/2), 
 m_4   \,\},\\ 
\chi^\mp_1 ~&=&~ \{\, (m_1, m_{23})^\pm, \mp  (m_{12} +m_{34}/2), m_{34}   \,\},\nn\\ 
\chi^\mp_{2,1} ~&=&~ \{\, (m_{12}, m_{23})^\pm, \mp  (m_{12} +m_{34}/2), 
m_{24}+m_2   \,\} ,\nn\\ 
\chi^\mp_{2,2} ~&=&~ \{\,( m_1, m_{24})^\pm, \mp (m_{12}+ m_3/2), 
m_{3}   \,\},\nn\\ 
\chi^\mp_3 ~&=&~ \{\, (m_{12}, m_{24})^\pm,\mp  (m_{12}+m_3/2), 
m_{23}+m_2   \,\},\nn\\ 
\chi^\mp_{3,1} ~&=&~ \{\, ( m_{2}, m_{13})^\pm,\mp(m_2 + m_{34}/2 ,  
 m_{14}+m_{12}   \,\},\nn\\ 
\chi^\mp_{3,2} ~&=&~ \{\, (m_{13}, m_{2})^\pm, \mp(m_{13}+m_4/2), 
m_{24}+m_{23}   \,\} ,\nn\\ 
\chi^\mp_{4,1} ~&=&~ \{\, (m_{14},m_{2} )^\pm,\mp  (m_2+m_3/2), 
m_{13}+m_{12}   \,\},\nn\\ 
\chi^\mp_{4,2} ~&=&~ \{\, (m_{13}, m_{24})^\pm, \mp  (m_{12}+m_3/2), 
m_{2}+m_{23}    \,\},\nn\\ 
\chi^\mp_{4,3} ~&=&~ \{\, (m_{23}, m_{12}^\pm, \mp(m_{23}+m_4/2), 
m_{14}+m_{13}   \,\} ,\nn\\ 
\chi^\mp_{4,4} ~&=&~ \{\, (m_{14}, m_{2})^\pm, \mp(m_{13}+m_4/2), 
m_{24}+m_{23}    \,\} ,\nn\\ 
\chi^\mp_5 ~&=&~ \{\, (m_{23}, m_{14})^\pm,\mp  (m_2+m_3/2),  
 m_{13}+m_{12}   \,\},\nn\\ 
\chi^\mp_{5,1} ~&=&~ \{\, (m_{13}+m_2, m_{24})^\pm, \mp (m_{12}+m_3/2), 
 m_{3}   \,\} ,\nn\\ 
 \chi^\mp_{5,2} ~&=&~ \{\, (m_{14}, m_{23})^\pm, \mp (m_{12}+m_{34}/2),  
m_{24}+m_{2}   \,\} ,\nn\\ 
\chi^\mp_{5,3} ~&=&~ \{\, (m_{24}, m_{12})^\pm, \mp(m_{23}+m_4/2), 
m_{14}+m_{13}   \,\} ,\nn\\ 
\chi^\mp_{5,4} ~&=&~ \{\, (m_{23}, m_{1})^\pm, \mp(m_{23}+m_4/2), 
m_{14}+m_{13}+2m_2   \,\} ,\nn\\ 
\chi^\mp_{6,1} ~&=&~ \{\, (m_{23}, m_{14}+m_2)^\pm,  \mp  (m_3/2), 
m_{13}+m_{12}   \,\},\nn\\ 
\chi^\mp_{6,2} ~&=&~ \{\, (m_{24}, m_{13})^\pm, \mp  (m_2 +m_{34}/2), 
m_{14}+m_{12}   \,\},\nn\\ 
\chi^\mp_{6,3} ~&=&~ \{\, (m_{14+m_2}, m_{23})^\pm, \mp (m_{12}+m_{34}/2), 
m_{34}   \,\} ,\nn\\ 
\chi^\mp_{6,4} ~&=&~ \{\, (m_{13}+m_2, m_{14})^\pm, \mp  (m_2 +m_{3}/2), 
m_{3}   \,\} ,\nn\\ 
\chi^\mp_{6,5} ~&=&~ \{\, (m_{24}, m_{1})^\pm, \mp (m_{23}+m_{24})/2, 
m_{14}+m_{13}+2m_2   \,\},\nn\\ 
\chi^\mp_{6,6} ~&=&~ \{\, (m_{2}, m_{1})^\pm, \mp (m_{2}+m_{34}/2), 
m_{14}+m_{13} +m_{23}+m_2  \,\} ,\nn\\ 
\chi^\mp_7 ~&=&~ \{\, (m_{24}, m_{13}+m_2)^\pm,\mp  (m_{34}/2), 
m_{14}+m_{12}   \,\},\nn\\ 
\chi^\mp_{7,1} ~&=&~ \{\, (m_{13}, m_{14}+m_2)^\pm, \mp  (m_{3}/2), 
m_{23}+m_{2}   \,\} ,\nn\\ 
\chi^\mp_{7,2} ~&=&~ \{\, (m_{2}, m_{14}+m_{23})^\pm, \pm  (m_{3}/2), 
m_{13}+m_{12}   \,\},\nn\\ 
\chi^\mp_{7,3} ~&=&~ \{\, (m_{24}, m_{1})^\pm, \mp (m_{2}+m_{34}/2), 
m_{14}+m_{13} +m_{23}+m_2  \,\} ,\nn\\ 
\chi^\mp_{7,4} ~&=&~ \{\, (m_{14}+m_{23},m_2)^\pm, \mp(m_{13}+m_4/2), 
m_{4}   \,\} ,\nn\\ 
\chi^\mp_{7,5} ~&=&~ \{\, (m_{14}+m_2, m_{13})^\pm, \mp (m_{2}+m_{34}/2), 
m_{34}   \,\},\nn\\ 
\chi^\mp_{7,6} ~&=&~ \{\, (m_{2}, m_{1})^\pm, \mp (m_{2}+m_{34}/2), 
2m_{14}+m_{23}+m_2   \,\},\nn\\ 
 \chi^\mp_{8,1} ~&=&~ \{\, (m_{24}, m_{13}+m_2)^\pm,\mp  (m_4/2),  
m_{13}+m_{14}   \,\},\nn\\ 
\chi^\mp_{8,2} ~&=&~ \{\, (m_{14}, m_{13}+m_2)^\pm,\mp  (m_{34}/2),  
 m_{24}+m_{2}   \,\},\nn\\ 
 \chi^\mp_{8,3} ~&=&~ \{\, (m_{24}, m_{12})^\pm, \mp  (m_{34}/2), 
m_{14}+m_{13} +m_{23}+m_2    \,\} ,\nn\\ 
\chi^\mp_{8,4} ~&=&~ \{\, (m_{23}, m_{1})^\pm, \mp (m_{2}+m_{3}/2), 
2m_{14}+m_{23}+m_2    \,\} ,\nn\\ 
\chi^\mp_{8,5} ~&=&~ \{\, (m_{14}+m_{23}, m_{12})^\pm, \mp (m_{23}+m_{4}/2), 
m_{4}   \,\} ,\nn\\ 
\chi^\mp_{8,6} ~&=&~ \{\, (m_{2}, m_{14}+m_{23})^\pm, \pm m_{34}/2, 
 m_{14}+m_{12}   \,\} ,\nn\\ 
\chi^\mp_{8,7} ~&=&~ \{\, (m_{12}, m_{14}+m_{23})^\pm, \pm m_{3}/2, 
m_{23}+m_{2}   \,\} ,\nn\\ 
\nn\eea

\eqnn{tabla}
\chi^\mp_9 ~&=&~ \{\, (m_{14}, m_{13}+m_2)^\pm, \mp  (m_4/2), 
m_{24}+m_{23}   \,\},\nn\\ 
\chi^\mp_{9,1} ~&=&~ \{\, (m_{24}, m_{13})^\pm, \mp  (m_4/2), 
m_{14}+m_{13} + 2m_2    \,\} ,\nn\\ 
 \chi^\mp_{9,2} ~&=&~ \{\, (m_{14}+m_{23}+m_2, m_{1})^\pm, \mp (m_{23}+m_{4}/2), 
 m_{4}   \,\} ,\nn\\ 
\chi^\mp_{9,3} ~&=&~ \{\, (m_{23}, m_{14}+m_2)^\pm, \pm m_{4}/2, 
 m_{14}+m_{13}   \,\},\nn\\ 
 \chi^\mp_{9,4} ~&=&~ \{\, (m_{14}, m_2)^\pm, \mp  (m_{34}/2), 
 m_{14}+m_{13} +m_{23}+m_2   \,\} ,\nn\\ 
\chi^\mp_{9,5} ~&=&~ \{\, (m_{1}, m_{14}+m_{23}+m_2)^\pm, \pm (m_{12}+m_{3}/2), 
 m_{3}   \,\} ,\nn\\ 
 \chi^\mp_{9,6} ~&=&~ \{\, (m_{12}, m_{14}+m_{23})^\pm , \pm m_{34}/2,  
 m_{24}+m_{2}   \,\} ,\nn\\ 
 \chi^\mp_{9,7} ~&=&~ \{\, (m_{23}, m_{12})^\pm, \mp  (m_{3}/2), 
 2m_{14}+m_{23}+m_2   \,\} ,\nn\\ 
  \chi_{10,1}^\mp ~&=&~ \{\, (m_{14}+m_2, m_{13})^\pm, \mp m_4/2, 
m_{23}+m_{24}   \,\}, \nn\\ 
\chi_{10,2}^\mp ~&=&~ \{\, (m_{13}, m_2)^\pm, \mp m_3/2, 
2m_{14}+m_{23}+m_2   \,\}, \nn\\   
\chi_{10,3}^\mp ~&=&~ \{\, (m_{14}, m_{23})^\pm, \mp m_4/2, 
m_{14}+m_{13}+2m_2   \,\},\nn\\ 
\chi_{10,4}^\mp ~&=&~ \{\, (m_{1},m_{14}+m_{23}+m_2)^\pm, \pm(m_{2}+m_{34}/2),  
m_{34}   \,\} 
\nn\eea
 The pairs ~$\L^\pm (\chi^\pm)$~ are symmetrically placed w.r.t. to the bullet in the middle of the figure.

\section{Concluding remarks}

\nt {\bf Remark 1:}~~\\
The integral intertwining KS operators act between the  spaces ~$\cc_{\chi^\mp}$~ in opposite directions:
\eqn{ksks} G^+_{KS} ~:~ \cc_{\chi^-} \llr \cc_{\chi^+}\ , \qquad
G^-_{KS} ~:~ \cc_{\chi^+} \llr \cc_{\chi^-} \ee

\nt {\bf Remark 2:}~~  \\
The  positive integers ~$\{ m_1,m_2,m_3,m_4\}$~ parametrize the finite-dimensional nonunitary irreps of ~$F'_4$~ (also
the unitary finite-dimensional   irreps of the compact Lie algebra $f_4$).

\nt {\bf Remark 3:}~~ \\
We expect that the discrete series are contained in the representation ~$\chi^+_0$~ since it is dual to  ~$\chi^-_0$~
where are contained the finite-dimensional (non-unitary) irreps. Following the Harish-Chandra criterion we must check which ~$\cm$-non-compact entries are negative. We recall that  the ~$\cm$-compact entries are ~$m'_1,m'_2,m'_{12},m'_4$~, all other
are non-compact. It is easy to see that all the ~$\cm$-non-compact entries are negative.  The discrete series irrep with lowest possible conformal weight ~$d=7$~
happens naturally when ~$m_1=m_2=m_3=m_4=1$. It corresponds to the one-dimensional irrep contained in ~$\chi^-_0$.

\bigskip

\nt {\bf Acknowledgments.}~~
  The author has received partial support from  Bulgarian NSF Grant DN-18/1.



 \np

 \pagestyle{empty}

 \voffset -4cm



\begin{figure}[]
\begin{center}
\includegraphics[width=1.4\hsize]{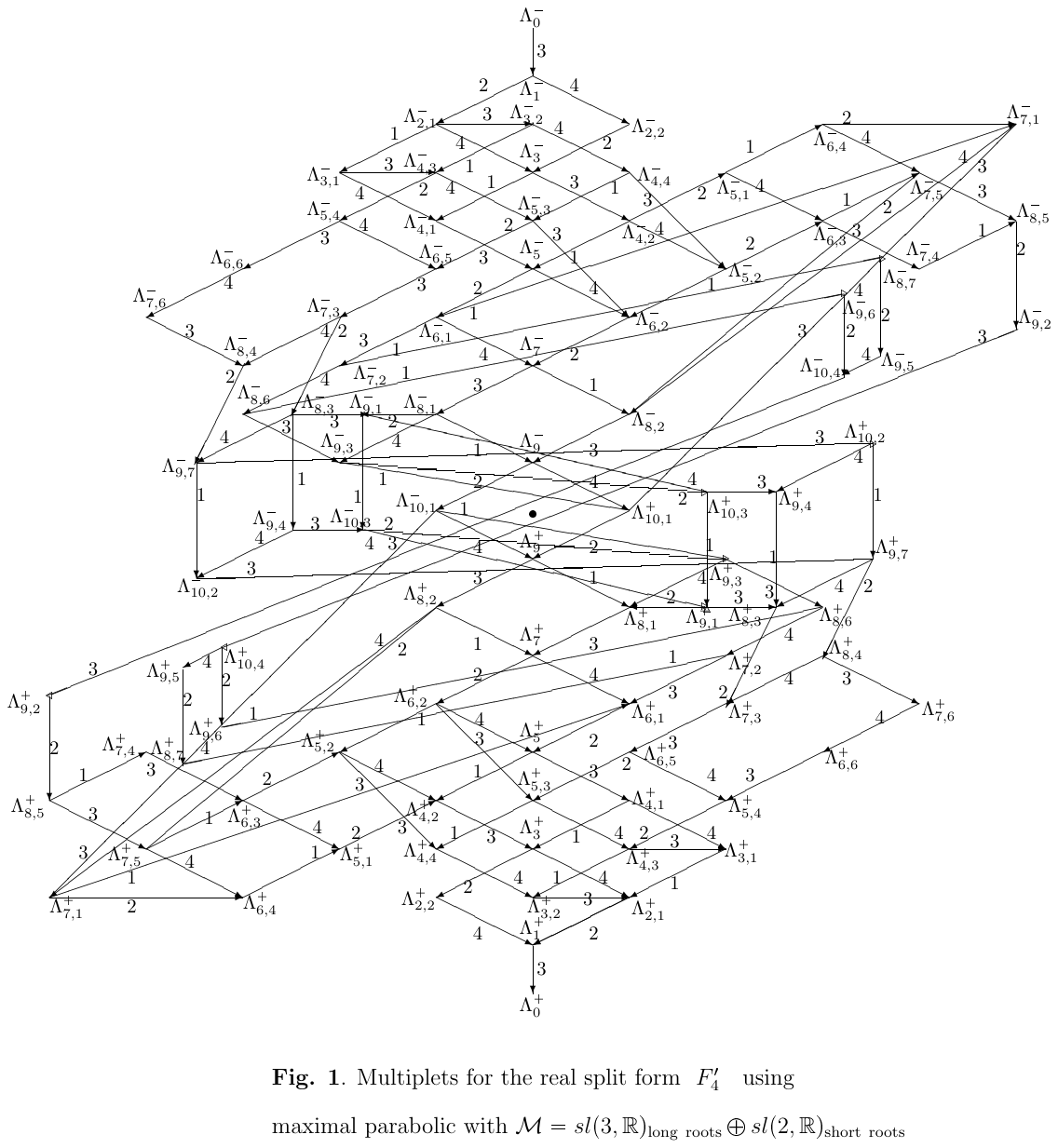} ~~~
\end{center}
\end{figure}




\newpage

\begin{thebibliography}{9}

\bigskip

\bibitem{Dob}V.K. Dobrev,
Rept. Math. Phys. {\bf 25}, 159-181 (1988) ; first as ICTP Trieste
preprint IC/86/393 (1986).

\bibitem{Dobinv}V.K. Dobrev, 
Rev. Math. Phys. {\bf 20} (2008) 407-449; hep-th/0702152.

\bibitem{Dobparab} V.K. Dobrev, 
J. High Energy Phys. 02 (2013) 015, arXiv:1208.0409.

\bibitem{Dobmul}V.K. Dobrev, Lett. Math. Phys. {\bf 9}, 205-211
(1985).

\bibitem{Dobk1} Vladimir K. Dobrev, {\it Invariant Differential Operators,
Volume 1: Noncompact Semisimple Lie Algebras and Groups}, De Gruyter
Studies in Mathematical Physics, vol. 35 (De Gruyter, Berlin, Boston,
2016).

\bibitem{Wein} S. Weinberg, {\it The quantum theory of fields} (vol 2)  (Cambridge University Press, 1996).

\bibitem{ITT}  M. Dubois-Violette and I. Todorov, 
Nucl. Phys. {\bf B938} (2019) 751-761,  arXiv:1808.08110 [hep-th].


 \bibitem{BGG} I.N. Bernstein, I.M. Gel'fand, S.I. Gel'fand,
"Structure of representations generated by highest weight vectors",
Funkts. Anal. Prilozh. {\bf 5}(1) 1-9(1971); English translation:
Funct. Anal. Appl. {\bf 5} 1-8 (1971).


\bibitem{Lan}R.P. Langlands, {\it On the classification of irreducible
representations of real algebraic groups}, Math. Surveys and
Monographs, Vol. 31 (AMS, 1988), first as IAS Princeton preprint
(1973).

\bibitem{Zhea}D.P. Zhelobenko, {\it Harmonic Analysis on Semisimple
Complex Lie Groups}, (Moscow, Nauka, 1974, in Russian).

\bibitem{KnZu}A.W. Knapp and G.J. Zuckerman, ``Classification theorems
for representations of semisimple groups'',
 in: Lecture Notes in Math., Vol. 587 (Springer, Berlin,
1977) pp. 138-159; 

\bibitem{DMPPT} V.K. Dobrev, G. Mack, V.B. Petkova, S.G. Petrova and I.T.
Todorov, {\it Harmonic Analysis on the n-Dimensional Lorentz Group
and Its Applications to Conformal Quantum Field Theory}, Lecture
Notes in Physics, No 63, 280 pages (Springer
Verlag, Berlin-Heidelberg-New York, 1977).




\bibitem{Knapp} A.W. Knapp,  {\it  Representation Theory of Semisimple
Groups (An Overview Based on Examples.)}, (Princeton Univ. Press 1986).

\bibitem{Dobf}V.K. Dobrev, in preparation.

\end{thebibliography}
\end{document}